# Vision Transformers with Autoencoders and Explainable AI for Cancer Patient Risk Stratification Using Whole Slide Imaging

Ahmad Hussein
Ahmad.Hussein@Student.UTS.edu.au, Mukesh Prasad
Mukesh.Prasad@uts.edu.au, Ali Anaissi
ali.anaissi@sydney.edu.au, and Ali Braytee
Ali.Braytee@uts.edu.au

University of Technology Sydney

**Abstract.** Cancer remains one of the leading causes of mortality worldwide, necessitating accurate diagnosis and prognosis. Whole Slide Imaging (WSI) has become an integral part of clinical workflows with advancements in digital pathology. While various studies have utilized WSIs, their extracted features may not fully capture the most relevant pathological information, and their lack of interpretability limits clinical adoption. In this paper, we propose PATH-X, a framework that integrates ViT and Autoencoders with SHAP (Shapley Additive Explanations) to enhance model explainability for patient stratification and risk prediction using WSIs from The Cancer Genome Atlas (TCGA). A representative image slice is selected from each WSI, and numerical feature embeddings are extracted using Google's pre-trained ViT. These features are then compressed via an autoencoder and used for unsupervised clustering and classification tasks. Kaplan-Meier survival analysis evaluates risk stratification into two and three risk groups. SHAP is applied to identify key contributing features, which are mapped onto histopathological slices to provide spatial context. PATH-X demonstrates strong performance in breast and glioma cancers, where a sufficient number of WSIs enabled robust stratification. However, performance in lung cancer was limited due to data availability, emphasizing the need for larger datasets to enhance model reliability and clinical applicability.

## 1 Introduction

Whole-slide imaging (WSI) has revolutionized digital pathology by enabling the digitization of entire histology slides, facilitating remote consultations and improving communication among healthcare providers, thereby making specialized care more accessible [6,7]. The integration of computational methods with WSI has enhanced the analysis and interpretation of histopathological slides, contributing to more precise and data-driven clinical decisions. Machine learning (ML) techniques have played a crucial role in computational pathology, offering automated solutions for tumor segmentation, classification, and prognostic modeling [12,1]. Various deep learning (DL)-based frameworks have been developed to process histopathological images, extracting meaningful features for cancer diagnosis and outcome prediction [6,7]. Pathomics focuses on extracting high-dimensional features from WSIs and converting them into compact representations for clinical decision-making [4]. Studies integrating deep learning with WSI and pathomics have demonstrated improved tumor detection and classification, supporting AI-driven pathology applications [14]. Despite these advancements, several research gaps remain in the field. Many existing deep learning models, particularly those leveraging convolutional neural networks (CNNs), lack interpretability, making their clinical adoption challenging. While some studies have utilized WSIs in their frameworks, their extracted features may not fully capture the most relevant pathological information, and their lack of interpretability limits clinical adoption. To address these gaps, we propose PATH-X, a framework that integrates explainable AI (xAI) techniques with ViT-extracted pathomic features. Our study evaluates the effectiveness of pathomics alone in risk stratification across multiple cancers.

The practical relevance of our proposed framework is evident, as it enables early detection of specific risk groups, helping clinicians make informed treatment decisions at different stages. Our approach enhances patient stratification in precision oncology. Furthermore, employing SHAP improves interpretability by highlighting key image features contributing to risk assessment. PATH-X was validated on BRCA, GLIOMA, and LUAD, demonstrating its potential for Pathomics feature extraction and multimodal integration. Our contributions are :

- We propose a framework for WSI-driven patient stratification that integrates an autoencoder with Google's pre-trained ViT to generate compact and meaningful feature representations.
- We integrate SHAP for enhanced model interpretability, ensuring explainable AI-driven decision-making.
- We run extensive experiments on three cancer types to showcase the effectivness of PATH-X in Pathomics analysis.

## 2   Related Work

Various studies have explored deep learning-based pathomics for cancer diagnosis, prognosis, and treatment response prediction. Nibid et al. [10] applied a deep learning-based pathomics approach to predict response to chemoradiotherapy in stage III NSCLC patients using 35 digitalized tissue slides. Five pre-trained CNN models (AlexNet, VGG, MobileNet, GoogLeNet, ResNet) were evaluated, with GoogLeNet achieving the best performance, correctly classifying responders (8/12) and non-responders (10/11), demonstrating high specificity (TNR: 90.1) and sensitivity (TPR: 0.75). Similarly, Li et al. [9] used WSI images of H&E-stained histological specimens from ESCC patients receiving PD-1 inhibitors, employing a pre-trained ViT model for patch-level feature extraction and an RNN-based patient-level predictor to construct an ESCC-pathomics signature (ESCC-PS). Their model, trained on 486,188 image patches, achieved 84.5% accuracy, effectively stratifying patients into three risk groups based on progression-free survival (PFS). Another study by Kim et al. [8] utilized WSIs from 256 melanoma patients to predict BRAF mutations using a CNN model and a pathomics-based pipeline. Their CNN approach identified tumor-rich areas (AUC=0.96) and predicted BRAF mutations (AUC=0.71), while pathomics-based analysis quantified nuclear morphology, revealing BRAF-mutated nuclei as larger and rounder than wild-type nuclei. A combined model integrating clinical data, deep learning, and pathomics improved predictive accuracy (AUC=0.89) for BRAF mutations. Furthermore, Williams et al. [6] reviewed various WSI-based deep learning approaches, including ensemble learning models to address class imbalance and high-dimensionality issues in tumor detection. Their study highlights the role of CNNs for tumor segmentation, weakly supervised learning for large-scale image analysis, patch-based aggregation for efficient feature learning, and visualization techniques like heatmaps and attention mechanisms to improve interpretability.

## 3 Methodology

### 3.1 Proposed Framework

We propose PATH-X, a deep learning framework for pathology-driven cancer prognosis and classification, comprising three key components: WSI slicing, feature extraction and dimensionality reduction, followed by feature evaluation, as shown in Fig. 1. Additionally, GradientSHAP enhances interpretability by localizing the most influential features on histological slices.GradientSHAP is an explainability method implemented in Captum, a model interpretability library for PyTorch. It approximates SHAP values using integrated gradients by computing the expectation of gradients over multiple noisy input perturbations and baseline references [13]. The framework extracts pathomics features from WSIs using Google's ViT model, generating high-dimensional representations. These features are then compressed via an autoencoder for efficiency, followed by clustering and classification for patient stratification.

### 3.2 Handling Whole Slide Images

Our method processes SVS files to extract $1024 \times 1024$ image slices, running on Google Colab's A100 GPU with cuCIM and CuPy for efficient computation. The "best slice" is selected using a composite scoring mechanism that integrates nuclei count (determined via color thresholding and connected component analysis), clarity (measured by Laplacian variance), and blank space.

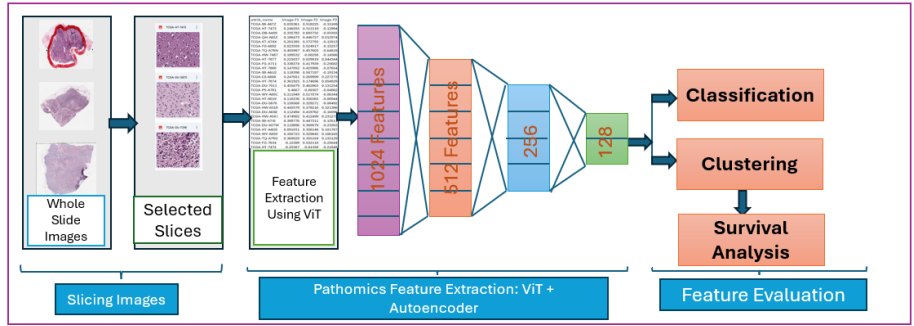

**Fig. 1.** PATH-X Framework: Image Slicing, Feature Extraction, Feature Evaluation

$$\text{score} = (\text{num\_nuclei} \times \text{clarity}) - \text{blank\_space} \tag{1}$$

Here, the number of detected nuclei (num_nuclei) is weighted by the clarity of the slice, while blank space is subtracted to penalize slices with large empty areas. This scoring mechanism ensures that only high-quality slices with well-preserved tissue structures are selected, facilitating accurate downstream analysis across various cancer types.

### 3.3   Pathomics Feature Extraction

To extract features from the resulting slices we use Google's Vision Transformer (ViT), introduced in the paper [5], an image $x \in \mathbb{R}^{H \times W \times C}$ is divided into $N$ patches of size $P \times P$, where:

$$N = \frac{H \times W}{P^2} \tag{2}$$

Each patch is flattened into a vector and linearly projected into D-dimensional embeddings, forming an input sequence to the transformer:

$$z_0^i = W_p x_p^i + E_p \tag{3}$$

where $W_p$ is the projection matrix, $x_p^i$ represents a patch, and $E_p$ is a learned positional embedding. These embeddings pass through multiple self-attention layers, following the Multi-Head Self-Attention (MHSA) mechanism:

$$\text{Attention}(Q, K, V) = \text{softmax}\left(\frac{QK^T}{\sqrt{d_k}}\right) V \tag{4}$$

where $Q, K, V$ are query, key, and value matrices derived from input embeddings. ViT processes images efficiently, achieving state-of-the-art performance on large datasets like ImageNet and demonstratingsuperior scalability compared to convolutional architectures.

### 3.4 Autoencoders

Autoencoders have been widely applied in dimensionality reduction, anomaly detection, and feature extraction [2]. Our employed autoencoder utilizes an encoder, decoder, loss function, and optimizer to effectively compress high-dimensional image features into a lower-dimensional latent representation. The encoder function transforms the 1024-dimensional input features into a 128-dimensional latent representation using fully connected layers with ReLU activation:

$$z = f_{\text{encoder}}(x) = \sigma(W_3 \sigma(W_2 \sigma(W_1 x + b_1) + b_2) + b_3) \tag{5}$$

where $W_1, W_2, W_3$ are the weight matrices for each dense layer, and $b_1, b_2, b_3$ are the corresponding biases. The decoder function reconstructs the original feature space from the latent representation, employing a similar multi-layer transformation with a sigmoid activation function:

$$\hat{x} = f_{\text{decoder}}(z) = \sigma(W'_3 \sigma(W'_2 \sigma(W'_1 z + b'_1) + b'_2) + b'_3) \tag{6}$$

where $W'_1, W'_2, W'_3$ and $b'_1, b'_2, b'_3$ are the weights and biases of the decoder layers. The sigmoid activation function ensures that values remain between 0 and 1, enhancing numerical stability:

$$\sigma(x) = \frac{1}{1 + e^{-x}} \tag{7}$$

To optimize reconstruction quality, the model minimizes the Mean Absolute Error (MAE) loss, defined as:

$$\mathcal{L}_{\text{MAE}} = \frac{1}{n} \sum_{i=1}^{n} |x_i - \hat{x}_i| \tag{8}$$

where $n = 1024$ is the number of input features, and $x_i, \hat{x}_i$ represent the original and reconstructed feature values, respectively. The model is trained using the Adam optimizer with a learning rate of 0.001, updating parameters as follows:

$$\theta \leftarrow \theta - \eta \frac{\partial \mathcal{L}}{\partial \theta} \tag{9}$$

where $\eta = 0.001$ is the learning rate, $\mathcal{L}$ is the MAE loss, and $\theta$ represents the model parameters (weights and biases).

## 4  Experiments

For this study, clinical data for Breast, Glioma, and Lung were sourced from the LinkedOmics portal, a publicly available resource providing multi-omics data for all 32 TCGA cancer types and 10 Clinical Proteomics Tumor Analysis Consortium (CPTAC) cancer cohorts [11]. Additionally, biospecimen diagnostic slides for TCGA cases were obtained from the GDC Data Portal using R. The datasets include 1,065 breast cancer cases, 880 glioma cases, and 478 lung cancer cases. For each case, 1,024 numerical features were extracted using Google's ViT model.

These high-dimensional features were then compressed into 128 latent features using an autoencoder. Hierarchical clustering was performed for each cancer type, with groups stratified into two and three clusters, and visualized using t-SNE. Clinical data for each cancer type were utilized to conduct Kaplan-Meier survival analysis, assessing the prognostic significance of the stratified risk groups. Furthermore, classification tasks were conducted to classify breast cancer cases based on tumor purity (low vs. high) and to distinguish glioma subtypes (GBM vs LGG). These classifications were evaluated using multiple machine learning models, including logistic regression, KNN, SVM, MLP, and random forest. Performance was assessed using accuracy, weighted F1 score, and macro F1 score as evaluation metrics. Finally, GradientSHAP was employed to identify the top 10 ViT-extracted features contributing to the encoded representations for each cancer type, allowing for the mapping of these critical features onto their corresponding histopathological slices to enhance interpretability.

## 5 Results

### 5.1 Glioma

Fig. 4 presents the t-SNE plots, demonstrating well-separated clusters and highlighting meaningful risk stratification based on Glioma Pathomics features extracted using PATH-X. Hierarchical clustering was employed to stratify patients into two and three risk groups, and the corresponding Kaplan-Meier survival curves further validate this stratification by showing significant survival differences across risk groups. The extremely low p-values (Table 1) reinforce the robustness of our method in distinguishing patient outcomes. Fig. 2 illustrates the integration of GradientSHAP, which identifies the most influential ViT-extracted features contributing to the encoded representations. The top 10 SHAP-selected features are visualized on histopathological slices of cases with the highest and lowest survival rates, providing insights into feature attribution and their role in downstream survival and clustering tasks. As shown in Table 2, various classification models were applied to classify glioma subtypes (GBM vs. LGG). Among these, SVM and Random Forest achieved the highest accuracy (0.8220) and weighted F1-score (0.8215).

### 5.2 Breast Cancer

We applied PATH-X to 1,065 BRCA cases and performed hierarchical clustering to stratify patients into risk groups based on their encoded features, which were visualized using t-SNE (Fig.5). We examined both two-group and three-group stratifications, with the corresponding Kaplan-Meier survival curves showing a clear separation between risk groups. The results indicate statistically significant p-values (7.78e-34 for two groups and extremely low values for three groups see Fig. 1), suggesting that our approach effectively distinguishes risk groups with clinically relevant survival differences. Fig. 3 shows the top 10 SHAP-selected features and their corresponding locations on histopathological slices of patients with the highest and lowest survival rates. Clinical validation is still necessary to assess whether the identified features correspond to meaningful pathological patterns. Table 2 presents the classification results for tumor purity (low vs. high) in BRCA cases using various classification methods.

| Cancer Type | p-values |
|---|---|
| **Glioma (2 Risk Groups)** | $7.78 \times 10^{-34}$ (Low vs High) |
| **Glioma (3 Risk Groups)** | $1.40 \times 10^{-23}$ (Low vs Medium) |
| | $2.05 \times 10^{-60}$ (Low vs High) |
| | $1.01 \times 10^{-18}$ (Medium vs High) |
| **Breast Cancer (2 Risk Groups)** | $2.517 \times 10^{-5}$ (Low vs High) |
| **Breast Cancer (3 Risk Groups)** | $1.024 \times 10^{-5}$ (Low vs Medium) |
| | $6.625 \times 10^{-12}$ (Low vs High) |
| | $4.524 \times 10^{-3}$ (Medium vs High) |
| **Lung Cancer (2 Risk Groups)** | $8.066 \times 10^{-1}$ (Low vs High) |

**Table 1.** Survival analysis p-values for different cancer types and risk groups.

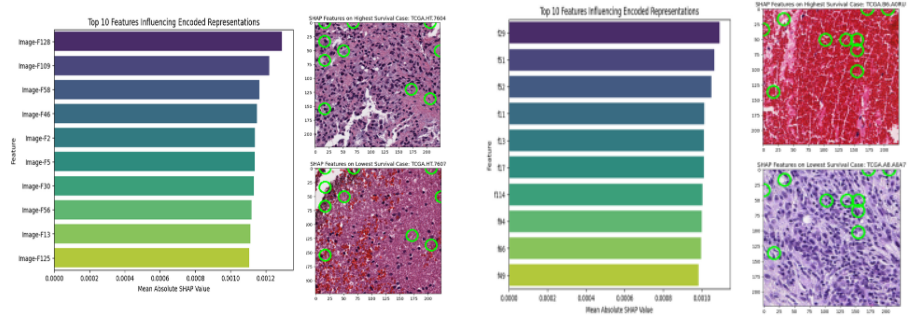

**Fig. 2.** Position of Top 10 SHAP Features on Highest and Lowest Survival for Glioma Cases

**Fig. 3.** Position of Top 10 SHAP Features on Highest and Lowest Survival for BRCA Cases

### 5.3 Lung Cancer

Our framework was applied to 478 lung cancer cases to stratify patients into two risk groups using hierarchical clustering based on encoded pathomics features. The t-SNE visualization showed moderate separation between the risk groups. However, Kaplan-Meier survival analysis resulted in a p-value of 0.8066, indicating no statistically significant survival difference between the groups. This suggests that the extracted ViT-based pathomics features alone may not be sufficient for robust risk stratification in lung cancer (LUAD). Such challenges could be attributed to the low samples and the inherent tumor heterogeneity.

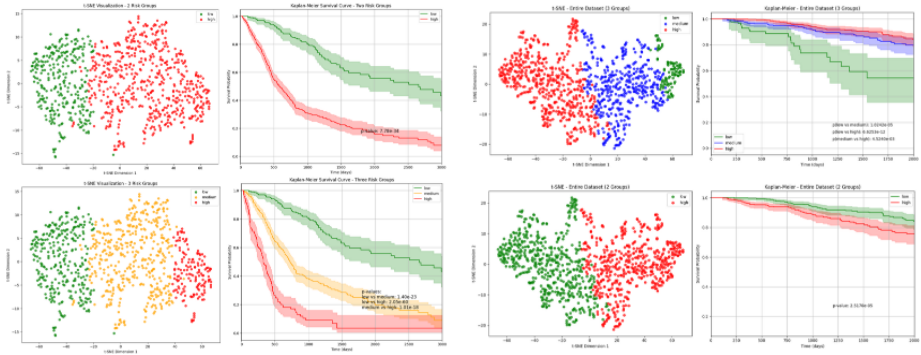

**Fig. 4.** t-SNE Visualization and Kaplan-Meier Survival Analysis for Glioma Risk Stratification

**Fig. 5.** t-SNE Visualization and Kaplan-Meier Survival Analysis for BRCA Risk Stratification

| Method | Glioma Subtype Classification | | | BRCA Tumor Purity Classification | | |
|---|---|---|---|---|---|---|
| | Accuracy | Macro F1 | Weighted F1 | Accuracy | Macro F1 | Weighted F1 |
| Logistic Regression | 0.8182 | 0.8155 | 0.8180 | 0.6151 | 0.5359 | 0.6030 |
| KNN | 0.7992 | 0.7927 | 0.7969 | 0.5994 | 0.5184 | 0.5875 |
| SVM | **0.8220** | **0.8188** | **0.8215** | 0.6119 | 0.5419 | 0.6046 |
| MLP | 0.7917 | 0.7888 | 0.7916 | 0.6057 | 0.5344 | 0.5982 |
| Random Forest | **0.8220** | **0.8188** | **0.8215** | **0.6215** | 0.5025 | 0.5877 |

**Table 2.** Classification Performance Comparison

### 5.4  Comparison with state-of-arts

Compared to the ESCC-PS framework [9], which utilizes a ViT-RNN architecture for Esophageal Squamous Cell Carcinoma (ESCC) patients receiving immunotherapy, PATH-X achieved significantly lower p-values, as demonstrated in Figures 5 and 4, while also incorporating explainability for a more interpretable model. Furthermore PATH-X outperformed the multiomics framework introduced in [3] in terms of P-value indicating stronger statistical significance in patient stratification. Despite utilizing only a single omic feature—pathomics. This suggests that histopathological features can serve as powerful biomarkers for risk stratification, even without incorporating genomic, transcriptomic, or epigenomic data. However, integrating additional omics layers could further refine patient subgroups and enhance predictive accuracy.

## 6  Conclusion

We proposed a framework utilizing ViT-extracted pathomics features for cancer patient stratification. Our method identified prognostically relevant features and achieved strong classification performance. SHAP analysis enabled interpretable feature selection, while feature mapping localized key features on histopathological slices. Challenges include the ViT model's fixed patching mechanism and the need for larger datasets. Despite these limitations, our approach demonstrates the potential of explainable pathomics analysis. Future improvements will focus on multi-omics integration and attention-based refinement. Expert validation will further enhance clinical relevance in precision oncology.